\documentclass[aps,twocolumn,groupedaddress]{revtex4}
\usepackage{graphicx}%

\begin{document}
\bibliographystyle{apsrev}


\title{An experience model for anyspeed motion}


\author{P. Fraundorf}
\email[]{pfraundorf@umsl.edu}
\affiliation{Physics \& Astronomy, U. Missouri-StL (63121), St. Louis, MO, USA}


\date{\today}

\begin{abstract}

	Simple airtrack simulations, like those now possible with web-based 
interactive 3D environments, can provide explorers of any age with 
experiential data sufficient to formulate their own models of motion.  
In particular, with a compressable spring, two gliders, a moving clock 
and two gate pairs with timers (pre and post collision), Newton's 
laws (or one's own version thereof) may emerge (or be tested) in the lab.  
At high speeds, one might find Minkowski's spacetime version of 
Pythagoras' theorem (the metric equation) in the data, along with 
``anyspeed" expressions for momentum and kinetic energy.  

\end{abstract}
\maketitle
\section{Introduction}

	Modeling workshop\cite{Hestenes87, Hestenes92} is a popular 
way to confront students with predictive challenges imposed by nature, 
and thereby allow them to discover their own (as well as 
conventional) strategies for meeting those challenges.  
Such ``hands on" approaches require first person experience with 
the phenomenon of interest if at all possible.  This is preferable 
in part because direct experience with nature is 
less dependent upon (i.e. is less mediated by) our choice of 
concepts for representing nature, than for example are: (i) tales about 
the experience of others, or (ii) conceptual explanations of how nature 
works.  Thus students can bring their own fresh eyes to the problem.

	Of course, there are many ``extreme physics" phenomena which 
are difficult, expensive, dangerous, or impossible to give 
students first hand access to.  Such phenomena include 
relativistic speeds, nanoscale structures, extreme space-time 
curvatures, wavefunction collapse, reciprocal space, and 
quantum tunneling.  For such phenomena, ``minimally mediated" 
experiences (sometimes but not always requiring simulations) are 
increasingly possible to arrange using virtual interfaces that 
require no specialized skills on the part of students (e.g. only 
pattern recognition and manipulative skills developed on the 
playground, or while playing video games).  Such experiences 
can give students a visceral appreciation of these phenomena, 
and perhaps also add motivation for acquiring the conceptual 
background (e.g. courses in calculus) likely required for 
a deeper understanding.

\section{The simulation}

\begin{figure}[tbp]
\includegraphics{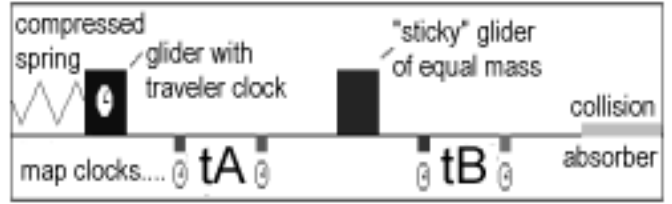}%
\caption{Schematic of the simulation.}
\label{Fig1}
\end{figure}

	Imagine the airtrack experiment illustrated in Fig \ref{Fig1}.  
Here a spring is compressed by distance $x_0$, and on release allowed to 
transfer kinetic energy $K_0$ to a glider.  The kinetic energy of the glider 
may be inferred from information on the spring's mass, displacement, and 
spring constant, or from information on the energy used to compress the spring.  Although compression distance is easier for introductory students to understand, 
use of kinetic energy (hence the amount of work done compressing a massless spring) 
as the independent variable is consistent with trends in physics education to introduce conservation of energy (e.g. \cite{Moore98}) as early as possible.  It 
is also consistent with 
the American Assocation for the Advancement of Science benchmarks\cite{benchmarks}, which propose energy as a `` major exception to the principle that students should understand ideas before being given labels for them''.  The simulation works 
in either case.

	The time interval elapsed as this glider then crosses gate-pair A is 
measured using lab-clocks ($t_A$) and a traveling-clock ($\tau_A$).  
This first glider then collides with, and sticks 
to, a 2nd glider. The time interval elapsed as the glider pair crosses 
gate-pair B is measured using lab-clocks ($t_B$), and the traveling clock 
on the first glider ($\tau_B$).  Choose conditions under which glider motion 
is essentially frictionless.

  The algorithms to run this simulation are relatively easy to derive.  
However, the focus in this note is on how students might get data from 
such experiments, and what to do with it after the fact, not to hedge 
student bets by providing equations before they have given the data a 
chance to speak for itself.

\section{Getting data from the simulation}

	Sample data from this experiment, e.g. elapsed times for various values of spring compression $x_0$ or launch energy $K_0$, are tabulated here.  The simulation was done twice, once with a rather flexible spring (Table \ref {Table1}), and the other with an extremely tight spring resulting in much higher glider speeds (Table \ref {Table2}).  In each case, the gate separations ($x_A$ and $x_B$) are 1 foot or 0.305 meters. The data contain some measurement errors in the times, although we've been careful to eliminate systematic errors, and to keep the random variations in measurement of the same time-interval at or about the 1 percent level.
\begin{table}
\caption{``Flexible spring data'' with some random measurement error, 
using $1$[kg] gliders, gate separations of $1$[foot] or $0.305$[m], and a spring constant of $1$[N/m].}
\begin{tabular}{rlcccc}
$x_0$[cm] & $K_0$[$\mu$J] & $t_A$[s] & $\tau_A$[s] & $t_B$[s] & $\tau_B$[s] \\
\hline
1 & 50 & 30.5 & 30.7 & 60.9 & 60.9 \\
2 & 200 & 15.4 & 15.2 & 30.6 & 30.5 \\
5 & 1250 & 6.15 & 6.13 & 12.2 & 12.3 \\
10 & 5000 & 3.08 & 3.03 & 6.13 & 6.07 \\
20 & 20000 & 1.53 & 1.54 & 3.03 & 3.07 \\
50 & 125000 & 0.61 & 0.60 & 1.23 & 1.22 \\
100 & 500000 & 0.31 & 0.30 & 0.61 & 0.60 \\
\end{tabular}
\label{Table1}
\end{table}
\begin{table}
\caption{``Tight spring data'' with some random measurement error, 
using $1$[kg] gliders, gate separations of $0.305$[m], and a spring constant 
of $1 \times 10^{18}$[N/m].}
\begin{tabular}{rlcccc}
$x_0$[cm] & $K_0$[TJ] & $t_A$[ns] & $\tau_A$[ns] & $t_B$[ns] & $\tau_B$[ns] \\
\hline
1 & $50$ & $30.4$ & 30.3 & 61.4 & 60.9 \\
2 & $200$ & $15.4$ & 15.1 & 30.8 & 30.3 \\
5 & $1250$ & $6.22$ & $6.11$ & 12.3 & 12.1 \\
10 & $5000$ & $3.17$ & 2.99 & 6.11 & 6.04 \\
20 & $20000$ & 1.78 & 1.44 & 3.08 & 2.88 \\
50 & $125000$ & 1.11 & 0.466 & 1.38 & 0.936 \\
100 & $500000$ & 1.03 & 0.158 & 1.06 & 0.314 \\
\end{tabular}
\label{Table2}
\end{table}

  Simulators with which the students can take data themselves also exist, or 
can be written.  We've developed a Visual Basic program which does this, and have more recently implemented a 3D simulator on the web 
using the beta-test version of Adobe's proposed Atmosphere environment server\cite{Fraundorf02a}.  The web simulator allows direct control (with a logarithmic slider) of the imparted kinetic energy (or the work done to compress a massless spring), rather than compression distance $x_0$.  The Adobe implementation can also serve as a meeting place for students, should they decide to take data as part of a team effort.  A Java applet for this simulation might also be useful, but is not available now.

\section{Discussion after the fact}

	The data generated by the simulation provides clues to natural rules of motion that are often explored in physics courses. Rather than tell students what others (including Aristotle, Galileo, Newton, and Einstein) might have proposed doing with such data, {\em modeling workshop} physics traditionally asks students: ``What patterns do you see in the data''.  One pattern at high speed that might be uncovered is a curious fact about the sum of squares of a traveler's ``speeds'' through space ($\Delta x / \Delta t$) {\em and time} ($c \Delta \tau / \Delta t$), where $c$ (lightspeed) is used to convert time elapsed on the clock of a traveler ($\tau$) into distance units.  See if any students happen to uncover this fact from the data itself.  

	A starting point for students is the classic modeling workshop strategy of plotting the data in various ways, to see if any simple (e.g straight line) relationships emerge.  Other questions to ask:  In the case of the flexible spring data set, does the relationship between lab-times ($t$) before and after collision seem to make sense?  How about the relationship between kinetic energy or spring compression, and the initial gate time $t_A$?  For the tight spring, to these also add: What's going on with the difference between lab and glider time-intervals (e.g. between $t_A$ and $\tau_A$)? 

	What concepts might be useful for cashing in on the insight from this experiment?  Aristotle might have said to pay attention to ``motives'' and speed.  Galileo in the 1500's might have said to consider accelerations as well\cite{Galileo62}.  Newton in the 1600's might have suggested the concept of momentum, and James Clerk Maxwell may have suggested considering energy\cite{Chandrasekhar95}.  Einstein in 1906 might have agreed, but then stressed the importance of specifying frame of reference when measuring positions and times.  More to the point:  What relationships and concepts does each student find useful or informative?  If they can determine this, then they will be able to claim as their very own as small part of our collective understanding of dynamics.  This alone qualifies them as participants in the development of better ways to understand such things downstream.
\bibliography{anyspeed.bib}

\end{document}